\documentclass[doublecol]{epl2} 

\usepackage{graphicx} 
\usepackage[usenames,dvipsnames]{xcolor}
\usepackage{amsmath}
\usepackage{amssymb}
\usepackage{color}
\usepackage{psfrag}
\usepackage{wasysym}
\usepackage{float}
\usepackage{epstopdf}
\usepackage[normalem]{ulem}

\newcommand{\taus}{\tau_{\mathrm{max}}}
\newcommand{\wo}{w_0}
\newcommand{\wid}{b}
\renewcommand{\S}{{\cal S}}
\newcommand{\dl}{\mbox{$\Delta L$}}
\newcommand{\Lbase}{L_{\mathrm{base}}}

\title{Dynamics of snapping beams and jumping poppers}
\author{A. Pandey\inst{1} \and D. E. Moulton\inst{2} \and D. Vella\inst{2} \and D. P. Holmes\inst{1}\thanks{E-mail:\email{dpholmes@vt.edu}}}
\shortauthor{A. Pandey \etal}

\institute{                    
  \inst{1} Department of Engineering Science and Mechanics, Virginia Polytechnic Institute and State University, Blacksburg, VA, 24061, USA\\
  \inst{2} Mathematical Institute, University of Oxford, Woodstock Rd, Oxford, OX2 6GG, UK }

\abstract{
We consider the dynamic snapping instability of elastic beams and shells. Using the Kirchhoff rod and F\"{o}ppl-von K\'{a}rm\'{a}n plate equations, we study the stability, deformation modes, and snap-through dynamics of an elastic arch with clamped boundaries and subject to a concentrated load. For parameters typical of everyday and technological applications of snapping, we show that the stretchability of the arch plays a critical role in determining not only the post-buckling mode of deformation but also the timescale of snapping and the frequency of the arch's vibrations about its final equilibrium state. We show that the growth rate of the snap-through instability and its subsequent ringing frequency can both be interpreted physically as the result of a sound wave in the material propagating over a distance comparable to the length of the arch. Finally, we extend our analysis of the ringing frequency of indented arches to understand the `pop' heard when  everted shell structures snap-through to their stable state. Remarkably, we find that not only are the scaling laws for the ringing frequencies in these two scenarios identical but also the respective prefactors are numerically close; this allows us to develop a master curve for the frequency of ringing in snapping beams and shells.
}

\begin{document}

\maketitle

\section{Introduction}

\begin{figure}[H]
\centering
\includegraphics[width=80mm]{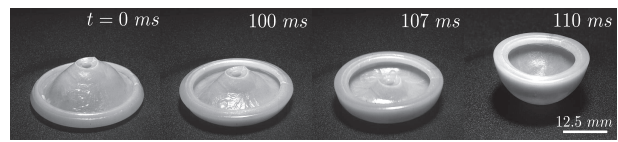}
\vspace{0mm}
\caption{Snap-through of a commercially available child's popper ($E\approx25\hspace{1mm}MPa$ and $\rho\approx1200\hspace{1mm}kg/m^3$) from a flat surface.}
\label{fig-1}
\end{figure}

 Arched structures have been used in architecture for over four thousand years~\cite{gordon78}, and the stability of these bistable structures has captivated scientists for over a century~\cite{hurlbrink1908,timoshenko1910}. The use of arches in engineering environments has traditionally placed a substantial focus on predicting the onset of loss of stability, and the subsequent unstable modes of deformation~\cite{fung1952,gjelsvik1962,pippard90,patricio98,pi2002,plaut2009}. More recently, the potential utility of controlling structural stability loss to take advantage of the rapid transition between two stable shapes has been the focus of study. Not only can a large deformation be achieved quickly, but frequently with little energetic cost.  Nature has long made use of such elastic instabilities for functionality, with the carnivorous waterwheel plant~\cite{poppinga2011} and Venus flytrap~\cite{forterre05,skotheim05} using their elaborate snapping leaves to rapidly capture their prey. Meanwhile, snapping shells have captivated children for decades in the form of bimetallic ``jumping" disks~\cite{wittrick1953} and rubber toy ``poppers"~\cite{lapp08,ucke09} that, having first been turned `inside-out' leap from a table with an audible pop as they return to their stable state (see fig.~\ref{fig-1}). These same principles have recently been used in the development of switches within MEMS devices~\cite{hsu2003,das2009}, biomedical valves~\cite{goncalves2003}, switchable optical devices~\cite{holmes07},  responsive hydrogels~\cite{xia2010}, and aerospace engineering \cite{Panesar:2012cj,Thill:2008uk}. 

Traditionally, studies of snapping have focussed on the conditions under which a shell or deformed beam can remain in equilibrium \cite{libai,forterre05}. Previous work on dynamic buckling tends to focus on how the critical load to induce snapping depends on the dynamics of loading \cite{Humphreys66,Pi13}. However,  in the design of advanced materials an understanding of the  dynamics of snapping itself is necessary to make full use of the snapping transition. In this Letter, we study the dynamics of snapping using a combination of experiments and theoretical calculations.

\section{Experiments}
Motivated by the `snap-through' of a popper shown in Fig.~\ref{fig-1}, we first study the snapping dynamics of a much simpler system: a two-dimensional elastic arch subject to a point load. We consider a shallow arch loaded by a point at its apex, and demonstrate the importance of the beam's `stretchability' on both the form of deformation and the snapping dynamics. In contrast to recent work that has focused on snapping induced by a fixed load \cite{Neukirch13, Pi13}, we consider the limit of `displacement control'. 
\begin{figure*}[t]
\begin{center}
\includegraphics[width=160mm]{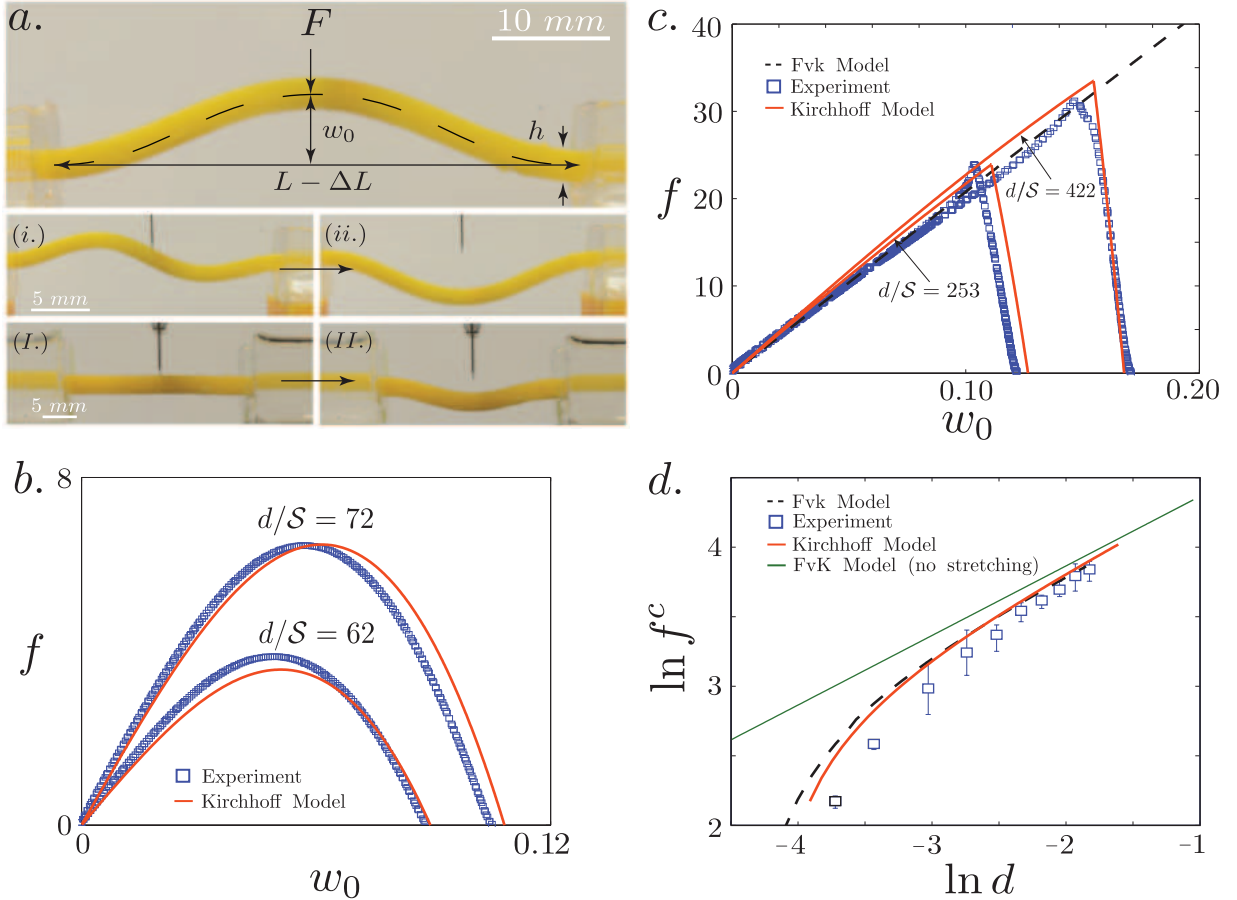}
\caption{Snap-through of a two-dimensional arch loaded along a line. a.) Experimental set-up showing a  Polyvinylsiloxane(PVS) arch with both ends clamped and loaded at its apex by a razor blade. (i) and  (ii) show the asymmetric deformation mode of a deep arch and the subsequent snap-through to the inverted stable configuration. (I) and  (II) show a shallow arch going through a flat mode of deformation and snapping. Force-displacement plots for arches that  b) remain symmetric throughout and c) first deform symmetrically but then asymmetrically. d.) Maximum force $f^c$ as a function of end--end compression $d$ for stretchability $\S_1=1.912\times10^{-4}$.}
\label{fig-2}
\vspace{-6mm}
\end{center}
\end{figure*}

Fig.~\ref{fig-2}a illustrates the setup considered here: a clamped beam of length $L$, thickness $h$, and width $\wid$  is compressed by an axial displacement $\Delta L$ so that it buckles into an arch. An imposed  displacement is then applied as a line load to the midpoint between the two clamped ends~\footnote{Polyvinylsiloxane (PVS, Elite 22, $Zhermack^{TM}$, $E=789.49\text{ }kPa$) beams were prepared with stretchability $\S_{1}=1.912\times10^{-4}$ and $\S_{2}=9.706\times10^{-4}$. The ends of the beam were clamped by embedding them within crosslinked  polydimethylsiloxane (PDMS) (Dow corning Sylgard $184^{TM}$). The indentation was performed in an Instron $5848$ Microtester and the force was measured by Interface ULC $0.5\text{ }N$ load cell.}. The initial height of the apex before indentation is $w_0$. Qualitatively speaking, we observe that for shallow arches, the beam remains symmetric about its center throughout the loading (fig.~\ref{fig-2}a.(I.)), while deep arches transition from a symmetric to an asymmetric shape well before snap-through occurs (fig.~\ref{fig-2}a.(i.)). In both cases, once the neutral axis of the beam is displaced to the mid-point of the clamped edges, global stability is lost, and the beam undergoes rapid snap-buckling to an inverted arch. To understand these observations, we begin by considering the motion before snapping, which we assume is quasi-static, before moving on to study the dynamics of snapping and ringing.

We  model the shape of the deformed beam using both the Kirchhoff equations for elastic rods and the F\"{o}ppl-von K\'{a}rm\'{a}n (FvK) equations \cite{audoly10}.  In the following discussion, we shall present the results of the FvK theory since this more easily allows for the identification of the important parameters and  the calculation of asymptotic results. However, numerical results from the Kirchhoff theory are used in comparisons with experiment since these account correctly for the effect of large displacements.

\section{FvK model}
In the case of small transverse displacement, the profile of the arch is denoted by $w(x,t)$, which satisfies the dimensionless dynamic beam equation \cite{audoly10}   
\begin{equation}
\frac{\partial^2 w}{\partial t^2}+\frac{\partial^4 w}{\partial x^4}+\tau^2\frac{\partial^2 w}{\partial x^2}=f\delta(x),\;\;-1/2<x<1/2.
\label{dyn_beam_nd}
\end{equation} Here, $\tau^2=TL^2/B$ is the dimensionless compressive force (with $T$ the dimensional axial load applied), $B=Eh^3\wid/12$ is the bending modulus (with $E$ the Young modulus), and $f=FL^2/B$ is the dimensionless indentation force applied at the center of the beam. Time is non-dimensionalized by $t_{\ast}=L^2\sqrt{12\rho_s h\wid/Eh^3\wid}=2\sqrt{3}L^2/ch$, where  $\rho_s$ is the density of the material and $c=\sqrt{E/\rho_s}$ is the  sound speed within the material.

The out-of-plane deflection $w$ is coupled to the in-plane horizontal displacement $u$ by Hooke's law relating the compressive stress to the higher-order von K\'{a}rm\'{a}n strain. In dimensionless terms, this reads
\begin{equation}
-\S \tau^2=\frac{\partial u}{\partial x}+\frac{1}{2}\left(\frac{\partial w}{\partial x}\right)^2, 
\label{axial_strain_nd}
\end{equation} where $$\S=\frac{B}{EhL^2b}=\frac{h^2}{12L^2}$$ is the `stretchability' of the beam and is  a measure of the relative importance of bending and stretching energies. The importance of stretchability has been discussed previously for vibrations about  equilibrium within the Kirchhoff formalism \cite{Neukirch2012} and in the context of wrinkling plates\cite{Hure2012}, though the analytical results and experiments presented here are, to our knowledge, new.

The clamped boundary conditions at the two ends of the arch are given by $w(\pm 1/2)={w}'(\pm 1/2)=0$ while the horizontal compression is imposed by the condition $u(\pm 1/2)=\mp d/2$, where $d=\dl/L$ is the dimensionless end-shortening.

\section{Quasi-static evolution}
 To study the deformation prior to snapping we assume that loading occurs quasistatically and consider time-independent solutions of \eqref{dyn_beam_nd}, i.e.~we neglect the $\partial^2_t w$ term. It is possible to solve the quasi-static equation for the shape of the indented beam analytically, the unknown indentation force $f$ being determined in terms of $w_0$ as part of the solution (for details, see supplementay information).
 
 Our analysis reveals that the form of deformation depends on the value of $d/\S$.  First, if $d/\S<4\pi^2$, the compressed, unbuckled beam is stable; neither buckling nor snapping occur as the in-plane compression does not induce Euler buckling in the beam. This can be understood in physical terms as follows: a compression $\dl$ corresponds to a strain $d$, a stress $Ed$ or a compressive force $Eh b d$ on the beam. However, the buckling load for a beam with clamped edges is well-known to be $F=4\pi^2 B/L^2=4\pi^2\S Ehb$; hence when $d<4\pi^2\S$ buckling cannot occur. In what follows the quantity $d-4\pi^2 \S$ frequently occurs and should be interpreted physically as the end-end compression that remains after some of the applied confinement of the beam has been accommodated by compressing the beam in response to the buckling load. 
 
 If $4\pi^2<d/\S<\taus^2\approx 80.76$ (with $\taus\approx8.99$ the solution of $\taus/2=\tan(\taus/2)$), the arch remains symmetrical as it deforms, ultimately returning to the compressed flat state $w=0$ (at which point snapping is observed experimentally). 
 
 For $d/\S>\taus^2\approx 80.76$, an asymmetric mode appears, in addition to the symmetric mode, once the indentation reaches a critical value or, equivalently, a critical force. Detailed calculations show that this asymmetric mode is energetically favourable whenever it exists. In the asymmetric mode, the force-displacement relation is linear, $f=-207.75w_0$, and the compressive force $\tau$ remains constant at $\tau=\taus\approx8.99$.  The FvK analysis also shows that the critical indentation force is given by  
\begin{equation}
f^{(c)}=-207.75w_0^{(c)}=-129.53(d-80.76\S)^{1/2}.
\label{eqn:critforce}
\end{equation}

While the theory could be tested by comparing theoretical and experimental beam shapes, a more rigorous test  is to compare the force-displacement relationship predicted theoretically with that measured experimentally for each of the three regimes discussed above.  In experiments, we have used beams with two different stretchabilities: $\S_1=1.912\times10^{-4}$ and  $\S_{2}=9.706\times10^{-4}$.  The regime of buckled beams with small compression ($4\pi^2<d/\S<\taus^2\approx 80.76$) is somewhat difficult to explore experimentally since the weight of the beam becomes important in this case. Therefore, to explore the force-displacement relation for arches in this regime we use the beam with relatively high stretchability ($\S_{2}=9.706\times10^{-4}$); the force-displacement plot in this case is shown in fig.~\ref{fig-2}b. For the regime of large compression, $d/\S>\taus^2\approx 80.76$, no such restriction applies; a comparison of theory and experiment in this regime, for two different values of $d/\S$, is shown in fig.~\ref{fig-2}c. 

Note that finite stretchability plays a crucial role in the picture outlined above. First, if stretchability is neglected, $\S=0$, then the family of beam shapes that return to a flat, compressed beam at $\wo=0$ (observed when $d<\taus^2\S$) cannot occur.   Second, even when $d/\S>\taus^2$  the critical transition force between symmetric and asymmetric deformations predicted in \eqref{eqn:critforce} is sensitive to the amount of stretchability, as is also seen experimentally (Fig~\ref{fig-2}d). It is in only in the high arch regime that stretchability becomes negligible \cite{Neukirch13}.

One final result from the study of the quasi-static indentation of an arch visible in fig.~\ref{fig-2}\emph{(b),(c)} is vital for our study of snapping: regardless of which deformation mode occurs, i.e.~independently of the precise value of $d/\S$, the indentation force and $w_0$ vanish together. This is also similar to what is observed in the confinement of a buckled arch in a shrinking box \cite{roman99,roman02}. If $4\pi^2<d/\S<\taus^2$, the solution at this point is the flat compressed beam, while for $d/\S>\taus^2$, the beam has the form of the antisymmetric, mode 2 Euler buckling.  In both cases, if the indentation were to continue, symmetry dictates that the indentation force would have to become negative. In the absence of any adhesion between indentor and arch, this is not possible and so no equilibrium solution is possible --- contact must be lost and the beam must then `snap-through' to the stable state, the inverted arch. This explains the experimental observation that snap-through occurs when the neutral  axis of the beam is displaced to the mid-point of the clamped edges. 

\section{Dynamics of snapping and ringing}

To understand the timescale of the snap-through, we perform a linear stability analysis of the beam at the point at which contact with the indentor is lost.  We set $f=0$ in eqn.\eqref{dyn_beam_nd} and look for solutions of the form $w(x,t)=w_{\alpha}(x)+\epsilon w_{p}(x)e^{\sigma t}$, where $w_{\alpha}(x)$ is the shape at the point of snapping ($\alpha=0,2$ depending on whether snapping occurs from the flat or asymmetric modes).  We also perturb the compressive force, $\tau=\tau_{\alpha}+\epsilon\tau_{p}e^{\sigma t}$. At leading order in $\epsilon$ we obtain an eigenvalue problem for the growth rate $\sigma$ with eigenfunction $w_{p}(x)$ satisfying
\begin{equation}
\frac{\mathrm{d}^4 w_{p}}{\mathrm{d} x^4}+\tau_{p}^2\frac{\mathrm{d}^2 w_{p}}{\mathrm{d} x^2}+\sigma^2 w_p=-2\tau_{\alpha}\tau_{p}\frac{\mathrm{d}^2 w_{\alpha}}{\mathrm{d} x^2},
\label{ode_per}
\end{equation}      
\begin{equation}
\int_{-1/2}^{1/2}\frac{\mathrm{d}w_{\alpha}}{\mathrm{d}x}\frac{\mathrm{d}w_p}{\mathrm{d}x}\mathrm{d}x=-2\S\tau_{\alpha}\tau_p,
\label{conf_per}
\end{equation}
along with boundary conditions $w_p(\pm1/2)={w}'_p(\pm1/2)=0$.
The eigenproblem \eqref{ode_per}-\eqref{conf_per} can be reduced to the solution of a transcendental equation for $\sigma$. Here, we summarize the results of this  analysis giving details in the supplementary information.

\begin{figure}
\begin{center}
\includegraphics[width=80mm]{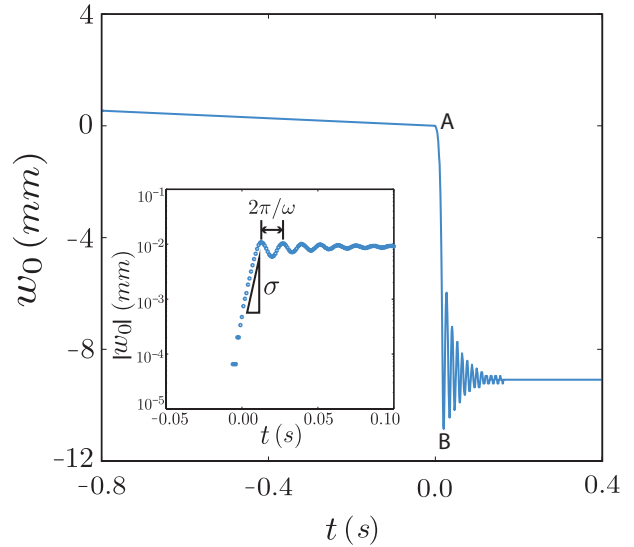}
\caption{Dynamics of arch snap-through. Main figure: Experimentally measured displacement of the central point as the arch snaps and vibrates. Inset: Semilog plot of displacement vs.~time, showing the growth rate $\sigma$ and vibration frequency $\omega$   after, snapping.}
\label{fig-3}
\vspace{-6mm}
\end{center}
\end{figure}

For $4\pi^2<d/\S<\taus^2$, for which the arch snaps from the compressed flat mode ($\alpha=0$), analytical insight may be obtained by considering compressions just large enough to obtain buckling, i.e.~$(d/\S)^{1/2}\gtrsim\tau_0=2\pi$. In this limit we find

\begin{equation}
\sigma\approx\frac{4\pi^{3/2}}{\sqrt{3}}\left[\left(\frac{d}{\S}\right)^{1/2}-2\pi\right]^{1/2}.
\label{sigma_st}
\end{equation}
For snap-through from the asymmetric mode ($\alpha=2$) there is a single eigenvalue of the system, $\sigma\approx24.113$, that is independent of the end-shortening $d$ and stretchability $\S$.  Thus, if we fix $\S$ and increase $d$ starting from $d=4\pi^2\S$, the growth rate increases monotonically from zero until $d=\taus^2\S$, at which point the growth rate plateaus and snap-through happens from the asymmetric mode. Numerical solutions of the Kirchhoff equations confirm a similar picture, though for $d/\S>\taus^2$ there is, in fact, a small dependence of $\sigma$ on $d/\S$, consistent with values reported in a related problem \cite{Neukirch13}.

\begin{figure}[t]
\begin{center}
\includegraphics[width=76mm]{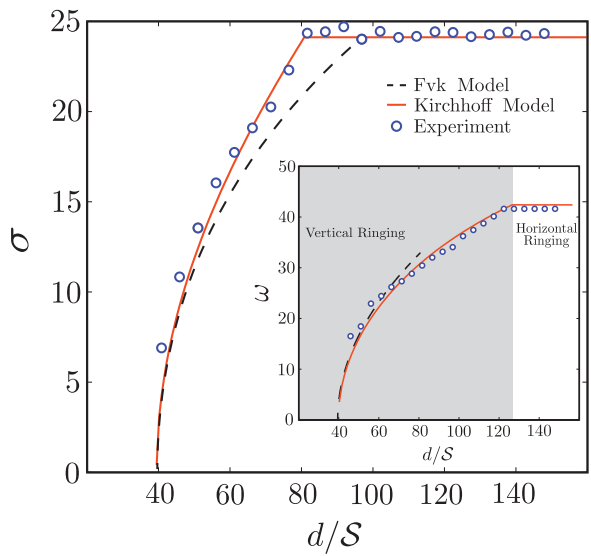}
\caption{Variation of growth rate $\sigma$ with rescaled lateral confinement, $d/\S$. Inset: The `ringing' frequency $\omega$ as a function of $d/\S$. The results of numerical simulations (solid curves) and experiments (points) are in good agreement with one another and with the prediction of the FvK model \eqref{sigma_st}, \eqref{eqn:ringfreq} (dashed curves) for $d/\S\approx4\pi^2$.}
\vspace{-6mm}
\label{fig-4}
\end{center}
\end{figure}

Experimentally, we tracked the motion of the center of the beam with a high speed camera (Photron FASTCAM APX RS, $@3000fps$) and performed image analysis with imageJ and custom MATLAB scripts to study the change in displacement with time. Fig.~\ref{fig-3} shows that as the center point reaches the base of the arch it rapidly moves from point A to point B, corresponding to the `snap-through'. The beam then vibrates like an under-damped oscillator about the `inverted' symmetrical shape before coming to rest. In the inset of fig.~\ref{fig-3} we plot the absolute value of the central deflection  between points A and B; this plot shows that $|w_0|$ grows approximately exponentially with time so that a growth rate can be measured. We non-dimensionalize the experimentally obtained growth rate using the time scale $t_*\sim L^2/ch$ that arises naturally from the dynamic beam equation; fig.~\ref{fig-4} shows that the time scale of snap-through observed experimentally agrees well with theoretical predictions. In particular, the growth rate $\sigma$ is strongly dependent on the degree of confinement for $4\pi^2<d/\S<\taus^2$, as predicted by Eqn.~\eqref{sigma_st}, but once $d/\S$ crosses the critical value $d/\S=\taus^2\approx80.76$, confinement plays a negligible role and $\sigma\approx24.113$.

Having snapped away from its unstable configuration,  the beam approaches the inverted arch state, which is a stable equilibrium. Since there is little dissipation in our system, the beam oscillates about this state. To understand the `ringing' frequency  $\omega=\sqrt{-\sigma^2}$ of this vibration we perform a linear stability analysis as before, but now with $w_\alpha=w_1$, the stable, first Euler buckling mode. Here, we again see a transition in the form of oscillation based on the quantity $d/\S$.  Starting at the critical value $d/\S=4\pi^2$, as $d/\S$ is increased the ringing frequency of the lowest mode of oscillation increases. The FvK model predicts that 
\begin{equation}
\omega\approx\frac{2^{3/2}\pi}{3^{1/2}}\left(\frac{d}{\S}-4\pi^2\right)^{1/2}
\label{eqn:ringfreq}
\end{equation} for $d/\S\gtrsim4\pi^2$. The Kirchhoff model confirms this result for $d/\S\approx4\pi^2$ and shows that $\omega$ continues to increase up to $\omega\approx42.38$ when $d/\S=122.4$. For $d/\S>122.4$, the frequency of the lowest mode is fixed at $\omega\approx42.38$, independent of  $d/\S$.  We also see a transition in the form of oscillation.  For $d/\S<122.4$, the mode of oscillation with the lowest frequency corresponds to a vertical oscillation, whereas for $d/\S>122.4$ the mode with lowest frequency also has a significant horizontal component. To compare this picture with experimental data the Fourier transform of the displacement data was taken to measure the frequency of the vibrations after snapping; fig.~\ref{fig-4} inset shows good agreement between the experiments and theory. The post-snap ringing timescale of the beam ($t_r^{(b)}$) for $d/\S>122.4$ is governed by the geometry and material properties of the arch: 
\begin{equation}
t_r^{(b)}=(2\pi/\omega)t_{\ast}\approx0.148t_{\ast}\approx0.513 L^2/ch.
\label{eqn:ringfplate}
\end{equation}

\section{Ringing of spherical shells}

The audible pop that accompanies the snapping of toys, such as jumping discs\cite{ucke09} and the popper shown in fig.~\ref{fig-1}, can be attributed to the ringing frequency.  It is therefore natural to compare the ringing timescale \eqref{eqn:ringfplate} with that for a spherical cap. The analogue of the beam length for a spherical cap is the base diameter, which we therefore denote by $\Lbase$. For an elastic spherical cap with radius of curvature $R$ and thickness $h$,  the analogue of the FvK equation \eqref{dyn_beam_nd} is  the dynamic Donnell--Mushtari--Vlasov equation \cite{soedel04}, which reads, in dimensional terms
\begin{equation}
\rho_sh\frac{\partial^2 w}{\partial t^2}+B\nabla^4w+\frac{Eh}{R^2}w=0,\quad 0\leq r\leq \Lbase/2,
\label{eqn:DMVdyn}
\end{equation} where $B=Eh^3/12(1-\nu^2)$ is the appropriate bending stiffness (with Poisson ratio $\nu$), $r$ is the radial coordinate, and we shall assume axisymmetry in what follows. Once a popper has leapt from a table, its edges are free and so the appropriate boundary conditions for \eqref{eqn:DMVdyn} are $\nabla^2w|_{r=\Lbase/2}=\upd/\upd r(\nabla^2w)|_{r=\Lbase/2}=0$. Performing a linear stability analysis along similar lines to those for the snapping beam (see supplementary information for more details) we obtain the snapping timescale
\begin{equation}
t_r^{(s)}=\frac{\pi}{2}\frac{\sqrt{1-\nu^2}}{\lambda^2} t_{\ast} \left(1+\frac{3}{4}\frac{1-\nu^2}{\lambda^4}\frac{\Lbase^4}{h^2R^2}\right)^{-1/2}
\label{eqn:ringfshell}
\end{equation} where $\lambda\approx3.196$ is an eigenvalue that is found numerically from a solvability condition.

\begin{figure}[t]
\begin{center}
\includegraphics[width=80mm]{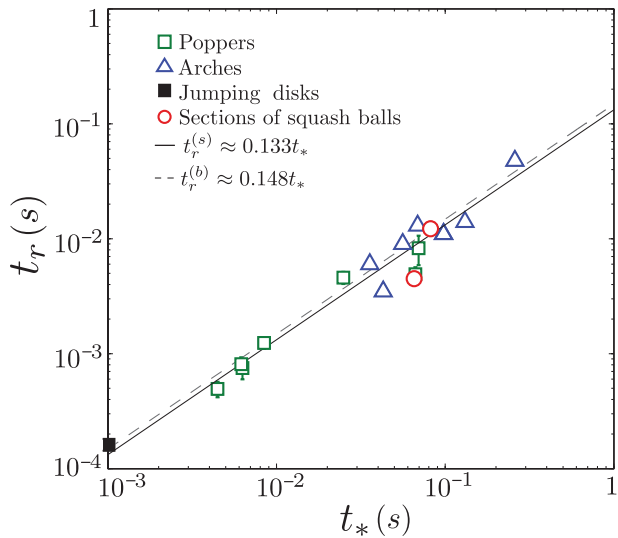}
\caption{The ringing timescale measured experimentally, $t_r$, scales with the time scale $t_*$ as predicted by our analysis for arches, $t_r^{(b)}\approx0.148t_{\ast}$ and for hemispherical shells $t_r^{(s)}\approx0.133t_{\ast}$.}
\vspace{-6mm}
\label{fig-5}
\end{center}
\end{figure}

For the majority of the shells of interest here, the term in parentheses in \eqref{eqn:ringfshell} is approximately unity and we find, assuming $\nu\approx1/2$, that $t_r^{(s)}\approx0.133t_\ast$. Remarkably, we find that this result is within $10\%$ of the corresponding result for the ringing frequency of arched beams in \eqref{eqn:ringfplate}. Given the quantitative robustness of the ringing timescale, it is natural to measure its value in a wide range of experiments combining the snapping shells that motivated our study with the carefully controlled snapping beams.  For spherical shells (commercially available toy poppers, bimetallic disks and sections of squash balls) we measure with a microphone the audible `pop' sound that they make during snapping  and extract the dimensional frequency, $\omega/t_\ast$ from this. The lengthscale used to estimate $t_\ast$ is the uncompressed length $L$ for arches and $\Lbase$ (the diameter of the spherical cap base) for hemispherical shells. Fig.~\ref{fig-5} shows plots of the experimentally measured ringing timescale, $t_r$, as a function of the characteristic timescale $t_\ast$; this confirms that the prediction of the linear stability analysis for both beams and shells, is in excellent agreement with experiments. 

\section{Conclusions}

We have studied the dynamic snapping of beams and shells. By first analyzing the quasi-static deformation of a point-loaded beam, we established the key role of stretchability in the small end-end compression regime, and showed that for $4\pi^2\S<d< 80.76\S$ the static deformation and dynamic snapping are fully 
symmetric, a regime that ceases to exist in the inextensible limit $S\to0$. However, we emphasize that this is not a generic feature of such systems; the loading also plays a crucial role since when indenting with a flat wall, the arch remains symmetric throughout the deformation independent of the stretchability \cite{roman99,roman02}. 

 Interestingly, our analysis and experiments showed that stretchability of the beam also plays a key role in determining the timescales that govern both snap-through and ringing. More generally, we found that the characteristic timescale in each case $\sim L^2/ch$, which may be interpreted physically as the timescale for the beam to `feel' its ends using sound waves of speed $c$, augmented by a geometric factor $L/h$. The augmenting factor, $L/h\sim \S^{-1/2}$, so that for a given stretchability $\S$ it is the time taken to `feel' the beam ends that limits the dynamics of motion. For shells, we showed that the timescale of ringing scales in the same way with the shell's properties and, moreover, with a very similar prefactor. This allowed us to present a universal description for the audible `pop' that is a distinctive feature of snapping both in everyday toys and laboratory experiments.

While our results have demonstrated that the ringing of arches and shells are quantitatively similar, many open questions remain to properly understand the dynamics of snapping structures. A particularly interesting open question concerns the effect of finite stretchability in determining the snapping growth rate for shells and whether such effects manifest themselves in terms of the symmetry (or asymmetry) of the snapping mode, as we have seen for snapping beams.

\acknowledgments

We are grateful to Howard Stone (Princeton) for his hospitality during the Oxford--Princeton Collaborative Workshop Initiative 2013, and the participants of that workshop, for useful suggestions in the completion of this work. D.V.~is partially supported by a Leverhulme Trust Research Fellowship. A.P.~and D.P.H.~acknowledge the financial support from NSF through CMMI-$1300860$.

\end{document}


\maketitle

This supplementary information gives details of the analytical results that can be obtained for the snapping problem by using the F\"{o}ppl-von-K\'{a}rm\'{a}n equations, and compares these with the results obtained by using the non-linear Kirchhoff equations. We begin by presenting the F\"{o}ppl-von-K\'{a}rm\'{a}n (FvK) results along with an outline of their derivation.  We also present a derivation of the ringing timescale of a spherical cap based on the Donnell-Mushtari-Vlasov equations.

\section{Problem setup}\label{SetupSec}

\begin{figure}
\centering
\includegraphics[width=12cm]{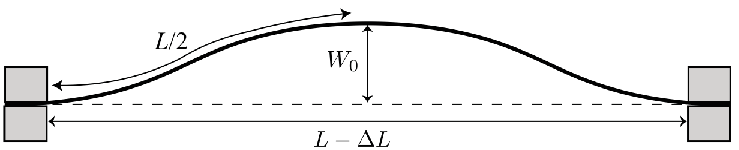}
\caption{The notation used for the geometry of a beam of length $L$, which is clamped at its ends.}
\label{fig:setup}
\end{figure}

The setup we consider is shown in figure \ref{fig:setup}. The geometrical parameters of the beam are its length $L$, thickness $h$ and width $\wid$ into the page; the height of the center of the beam (measured relative to the clamped ends) is $\wo=w(0)$. The mechanical properties of the beam are its Young's modulus, $E$, Poisson ratio $\nu$ and density $\rs$. The modulus $E$ often appears in combination with the thickness $h$ and beam width $\wid$ to give the bending stiffness
\beq
B=\frac{Eh^3\wid}{12}.
\label{Bending}
\eeq Note that because we are considering a thin strip rather than an infinitely wide plate there is no factor of $1-\nu^2$ in the denominator here \cite[see][for example]{audoly10}.

Under the assumptions of small slopes made in the derivation of the FvK equations, the following dynamic beam equation describes the profile $w(x,t)$ of a beam in response to a point force $F$ acting at the origin: 
\beq
\rs \wid h{\partial^2w\over \partial t^2}+B{\partial^4w\over \partial x^4}+T{\partial^2w\over \partial x^2}=F\delta(x).
\label{eqn:beameqndim}
\eeq Here $T$ is the compressive force applied, which is related to the strain in the $x$ direction, $\epsilon_{xx}$, by Hooke's law $\epsilon_{xx}=-T/Eh\wid$. The strain $\epsilon_{xx}$ is also related to the horizontal and vertical displacements $u$ and $w$, respectively, by
\beq
\epsilon_{xx}=\pd{u}{x}+\tfrac{1}{2}\left(\pd{w}{x}\right)^2.
\eeq We therefore have that
\beq
-\frac{T}{Eh\wid}=\pd{u}{x}+\tfrac{1}{2}\left(\pd{w}{x}\right)^2.
\label{eqn:Hookedim}
\eeq
The two equations \eqref{eqn:beameqndim} and \eqref{eqn:Hookedim} are to be solved with the boundary conditions
\beq
w(\pm L/2)=w'(\pm L/2)=0,\quad u(\pm L/2)=\mp \dl/2.
\label{eqn:bcsdim}
\eeq
When the beam is in contact with the indentor, one can either prescribe a fixed load or a fixed displacement.  Here we adopt the latter approach;  we  impose the condition $w(0)=\wo$, which provides an additional condition that allows the required force $F$ to be determined.

To non-dimensionalize the problem, we scale lengths by the length of the beam, $L$, and introduce dimensionless variables $w'=w/L$, $x'=x/L$, etc.  Since the delta function has units $1/[L]$ it is natural then to introduce the dimensionless indentation force $f=F\times L^2/B$. Time is scaled by $t_*=L^2\left(\rs h\wid/B\right)^{1/2}$, i.e. we introduce $t'=t/t_*$.  Note that this time scale can be written as $t_*=2\sqrt{3}{L\over c}{L\over h}$, where $c=(E/\rho)^{1/2}$ is the speed of sound in the material.

Inserting these scalings into \eqref{eqn:beameqndim} and \eqref{eqn:Hookedim} (and dropping primes henceforth) we have
\beq
{\partial^2w\over \partial t^2}+{\partial^4w\over \partial x^4}+\tau^2{\partial^2w\over \partial x^2}=f\delta(x)
\label{eqn:beameqnnd}
\eeq
and
\beq
-\S \tau^2=\pd{u}{x}+\tfrac{1}{2}\left(\pd{w}{x}\right)^2
\label{eqn:Hookend}
\eeq where $\tau^2=TL^2/B$ is the dimensionless compression force and
\beq
\S=\frac{B}{EhL^2\wid}=\frac{h^2}{12L^2}
\label{eqn:Sdefn}
\eeq is the dimensionless `stretchability' for reasons that shall shortly become apparent. The boundary conditions \eqref{eqn:bcsdim} have the obvious forms, though it is worth noting that the condition on the displacement $u$ becomes
\beq
u(\pm 1/2)=\mp d/2
\label{eqn:ubcnd}
\eeq where $d=\dl/L$ is the ratio of  end shortening to the length of the beam.  

We note that upon integrating \eqref{eqn:Hookend} over $-1/2\leq x\leq 1/2$ and applying \eqref{eqn:ubcnd} we have
\beq
\int_{-1/2}^{1/2}\left(\pd{w}{x}\right)^2~\upd x=2\left(d-\S\tau^2\right).
\label{eqn:constraint}
\eeq Geometrically speaking, this equation shows how the amplitude of the beam deformation must be chosen to accommodate the imposed end-shortening $d$. However, the presence of a non-zero stretchability, $\S$, reduces the effective amount of end-shortening felt by the beam and hence reduces the amplitude of its deformation. The factor $\S$ therefore represents the propensity of the beam to compress/stretch in response to externally applied stresses.

\section{Statics}

To simplify the analysis, we begin by neglecting the time derivatives that appears in \eqref{eqn:beameqnnd}. This corresponds to the assumption that the loading of the arch occurs quasistatically. In this limit the partial derivatives all become ordinary $x$-derivatives and \eqref{eqn:beameqnnd} becomes
\beq
\df{w}{x}+\tau^2\dd{w}{x}=f\delta(x),
\label{eq:linbeam}
\eeq

Considering each of the intervals $-1/2\leq x<0$ and $0<x<1/2$ separately, we have 
\beq
\df{w}{x}+\tau^2\dd{w}{x}=0\;\text{ on }-1/2\leq x<0,\;\;\;0<x<1/2
\label{eq:linbeam}
\eeq
 subject to the boundary conditions
\beq
w(\pm1/2)=w'(\pm1/2)=0,\quad w(0)=\wo,\quad [w]_-^+=[w']_-^+=[w'']_-^+=0,\;\;[w''']_-^+=f
\eeq and the imposed confinement
\beq
2(d-\tau^2\S)=\int_{-1/2}^{1/2}\left(\d{w}{x}\right)^2~\id x.
\label{eq:intconst}
\eeq Here $[g]_-^+\equiv g(0^+)-g(0^-)$ denotes the discontinuity in the quantity $g$ across the indentation point $x=0$.

The general solution of \eqref{eq:linbeam} is given in the Appendix.  Here, it is enough to note that there are two classes of solutions: symmetric and asymmetric.  Also, as \eqref{eq:intconst} implies the requirement that $d-\tau^2\S\geq0$, the form of solution is different based on the quantity $d/\S$, which is fixed for any given experiment.  The initial buckling of the beam occurs in the symmetric state and corresponds to $\tau=2\pi$.  Thus, if $d/\S<4\pi^2$, there is no buckling, i.e. the compressed flat state is the only solution, and $d=4\pi^2\S$ defines the critical shortening required to give buckling.
  
In the asymmetric state, $\tau=\taus\approx8.99$ (the smallest solution of $\taus/2=\tan\taus/2$) is fixed independent of $w_0$ or $f$.  Hence if $4\pi^2<d/\S<\taus^2$, the asymmetric solution does not exist, and the symmetric mode must be observed throughout indentation, even until $w_0=0$, at which point we reach the second buckling mode of the beam and so $f=0$ also.  

In the case $d/\S>\taus^2$, both symmetric and asymmetric solutions can be present, depending on $w_0$.  For small indentations, $\tau\approx2\pi$ and only the symmetric solution exists.  As the indentation progresses, $w_0$ decreases to $\woc\approx 0.624(d-80.76\S)^{1/2}$ while $\tau$ increases to $\taus$.  Beyond this point both modes exist, but the asymmetric mode is energetically favourable.  Thus the shape transitions to the asymmetric mode at $w_0=\woc$ and remains in an asymmetric state as the midpoint is indented from $w_0=\woc$ to $w_0=0$.


In summary, if $d/\S\geq 80.76$ then we observe two different modes:

\begin{itemize}

\item the symmetric mode for $$0.62(d-80.76\S)^{1/2}\leq\wo\leq 2(d-4\pi^2\S)^{1/2}/\pi$$ 

\item the asymmetric mode for $$0\leq \wo\leq0.62(d-80.76\S)^{1/2}$$ 

\end{itemize}

If $4\pi^2\leq d/\S\leq 80.76$ we observe only the symmetric mode. The corresponding force laws are

\beq
f=\begin{cases}
 g[\tau(\wo)]~\wo, \quad \mbox{symmetric mode},\\
-207.75~\wo, \quad\mbox{asymmetric mode},
\label{eqn:forcelaw}
\end{cases}
\eeq where
\beq
g(\tau)=\frac{2\tau^3\sin\tau/2}{2-2\cos\tau/2-\tau/2\sin\tau/2}
\eeq and $\tau$ satisfies
\beq
\wo=(d-\tau^2\S)^{1/2}\frac{\tau\cos\tau/4-4\sin\tau/4}{\tau^{1/2}\left(2\tau+\tau\cos\tau/2-6\sin\tau/2\right)^{1/2}}.
\eeq It is interesting to note that the second part of the force law in \eqref{eqn:forcelaw} is universal and does not depend on the values of either $d$ or $\S$.

Finally, we note that the maximum force observed is that at the transition between symmetric and asymmetric modes, i.e. 
\beq
f^{(c)}=-207.75~\woc=-129.53(d-80.76\S)^{1/2}.
\label{eqn:peakforce}
\eeq 

\section{Dynamics}\label{Dyn_section}
We now want to explore the snap-through dynamics.  As we have established above, the form of equilibrium deformation depends on the quantity $d/\S$.  
If $d/\S<4\pi^2$, the flat compressed state is stable, and clearly there is no snapping.  For $4\pi^2<d/\S<\taus^2$, the solution corresponding to $w_0=f=0$ is the flat, compressed but unstable beam, whereas if $d/\S>\taus^2$, the $w_0=f=0$ solution is the unstable, antisymmetric  (mode 2) solution of an end-shortened beam with no applied force.  
%
In either case, the fact that the force vanishes at this point, combined with the fact that these solutions are unstable (to be shown subsequently), suggests that snap-through occurs at this point. Alternatively, from the analysis above we see immediately that if the indentation were to continue for $\wo<0$ then the vertical force would have to become negative: this corresponds to an attractive interaction between the indentor and the beam and, assuming no such adhesive attraction exists, is not possible. Therefore snap-through must occur.  

To explore this snap-through, we study the dynamic problem.  In both cases, the snap-through takes the system from an unstable state with $w_0=0$ to the stable, symmetric solution, i.e. the inverted form of the initially arched beam.   Since the applied force due to the indentor plays no role in the dynamics, we can examine the timescale of snap-through by investigating the linear stability of both of the states with $\wo=0$ and determining the growth rate of the instability, $\sigma$.  We can also investigate the ringing frequency of the beam once it reaches the inverted state via a linear stability analysis of the stable, symmetric arched beam.


\subsection{Linear stability analysis of snapping}\label{Section_lin_stab_FvK}
Once contact is lost with the indentor, the evolution of the beam shape during snapping is governed by \eqref{eqn:beameqnnd} with $f=0$, i.e.
\beq
{\partial^2w\over \partial t^2}+{\partial^4w\over \partial x^4}+\tau^2{\partial^2w\over \partial x^2}=0.
\eeq

We look for a solution of the form $w(x,t)=w_\alpha(x)+\epsilon w_p(x)e^{\sigma t}$, $\tau=\tau_\alpha+\epsilon \tau_p e^{\sigma t}$. Here, $w_\alpha(x)$ is the unperturbed shape ($\alpha=0,2$ depending on whether this is the mode $0$ or mode $2$ state) and $\tau_\alpha$ is the corresponding compression. $w_p(x)$ and $\tau_p$ are the perturbations to the shape and compression, respectively. At $O(\epsilon)$, we have the following ODE for $w_p(x)$
\beq
{\upd^4w_p\over \upd x^4}+\tau_\alpha^2{\upd^2w_p\over \upd x^2}+\sigma^2w_p=-2\tau_\alpha\tau_p \frac{\upd^2 w_\alpha}{\upd x^2},
\label{eqn:perturbed}
\eeq
which should satisfy the boundary conditions $w_p=w_p'=0$ at $x=\pm1/2$ as well as the length constraint given by expanding \eqref{eq:intconst} to $O(\epsilon)$
\beq
\int_{-1/2}^{1/2}\frac{\upd w_\alpha}{\upd x}\frac{\upd w_p}{\upd x}~\upd x=-2\S\tau_\alpha\tau_p.
\label{eqn:intconst_pert}
\eeq

It might reasonably be expected that the function $w_p(x)$ will be an even function of $x$ since it has to get the beam `close' to the symmetrical inverted beam shape. From this expectation, \eqref{eqn:intconst_pert} shows that in the case of snap-through from the (odd) mode 2 state, the integrand in \eqref{eqn:intconst_pert} is odd, and so the LHS is zero and thus $\tau_p$ must equal 0. A detailed calculation shows that this heuristic expectation is, in fact correct, and that $\tau_p=0$ to within numerical errors. We therefore make this assumption in the analysis that follows to clarify the presentation.  For snap-through from the flat mode 0 state (the compressed state), $w_0\equiv0$ and so we immediately have that $\tau_p=0$ in this case too. We shall therefore assume that $\tau_p=0$ in what follows; repeating the calculation without either this assumption or that of the evenness of $w_p$ confirms that both of these assumptions are in fact valid.

Seeking a symmetric solution of \eqref{eqn:perturbed} we find that the growth rate $\sigma$ must be such that
\beq
\lambda_+\tan\lambda_+=\lambda_-\tan\lambda_-
\label{eqn:lameqn}
\eeq where
\beq
\lambda_\pm^2=\tfrac{1}{8}\left[\tau_\alpha^2\pm(\tau_\alpha^4-4\sigma^2)^{1/2}\right].
\eeq

For snap-through from the mode 2 Euler buckling solution, i.e.~$\tau_\alpha=\taus$, we find that the system has a single eigenvalue, $\sigma\approx24.11$. We note that this value is independent of the end-shortening $d$ and stretchability $\S$.

For snap-through from the mode 0 compressed solution, the value of $\tau$ about which we are linearizing is that for which $w_0\equiv0$, i.e.~$\tau_\alpha=(d/\S)^{1/2}$. The value of $\sigma$ as a function of $d/\S$ can be determined numerically. However, for $\tau_\alpha\approx2\pi$, i.e. close to the critical compression required for arch buckling, we expect that the snap-through instability should disappear. By letting $(d/\S)^{1/2}=\tau_\alpha=2\pi+\epsilon$ with $\epsilon\ll1$ and expanding \eqref{eqn:lameqn} as a power series in $\epsilon$ we find that $\sigma\approx 4\pi^{3/2}/\sqrt{3}\epsilon^{1/2}$; we therefore have
\beq
\sigma\approx \frac{4\pi^{3/2}}{\sqrt{3}}\left[\left(\frac{d}{\S}\right)^{1/2}-2\pi\right]^{1/2}.\label{sigma_approx}
\eeq Thus we see that the growth rate of the zeroth mode does depend on the amount of end-shorterning $d$, increasing from 0 when $d/\S=4\pi^2$ and the compressed state is unstable.

\subsection{Linear stability analysis of ringing}

The previous analysis gives a characteristic time scale for the snap-through instability. To compute the frequency of the `ringing' that is observed after snap-through we must perform a similar analysis but for a perturbation about the first buckling mode, i.e.~we let $w(x,t)=w_1(x)+\epsilon w_p(x)e^{\sigma t}$, $\tau=\tau_1+\epsilon \tau_p e^{\sigma t}$ where
\beq
w_1(x)=\alpha(1+\cos2\pi x)
\eeq $\alpha=-\left(\frac{d}{\pi^2}-4\S\right)^{1/2}$ and $\tau_1=2 \pi$.

We find that
\begin{eqnarray*}
\frac{w_p(x)}{\alpha}=\frac{16\pi^3}{\sigma^2}\tau_p\cos2\pi x&+&A\cosh\bigl[\lambda_-(x+1/2)\bigr]+B\sinh\bigl[\lambda_-(x+1/2)\bigr]\\&+&C\cos\bigl[\lambda_+(x+1/2)\bigr]+D\sin\bigl[\lambda_+(x+1/2)\bigr]
\end{eqnarray*} where
\beq
\lambda_\pm=\left[\left(4\pi^4-\sigma^2\right)^{1/2}\pm 2\pi^2\right]^{1/2}.
\eeq The unknowns $\tau_p, \sigma$, $A,B,C$ and $D$ are determined as an eigenproblem based on the boundary conditions
\beq
w_p(\pm1/2)=w_p'(\pm1/2)=0
\eeq and integrated Hooke's law
\beq
2\tau_p\S-\alpha\int_{-1/2}^{1/2}\frac{\upd w_p}{\upd x}\sin2\pi x~\upd x=0
\eeq with eigenvalue $\sigma^2$. This eigenproblem can easily be solved numerically.

For all values of $d/\S>4\pi^2$ we find the eigenvalue $\sigma^2\approx-1968.05$, i.e.~the solution is stable, and the perturbation gives an oscillation with dimensionless frequency $\omega=\sqrt{-\sigma^2}\approx44.36$. In this case, the eigenfunction $w_p(x)$ is an odd function of $x$ and so the combined behaviour gives the illusion of a horizontally oscillating mode, even though the perturbation is only an up-down oscillation.

Another eigenvalue exists, which does depend on the value of $d/\S$. The eigensolution associated with this eigenvalue is even and corresponds to the up-down oscillation that is observed experimentally. For $d/\S-4\pi^2\ll1$ we find that this mode has
\beq
\sigma^2\approx-\frac{8\pi^2}{3}\left(\frac{d}{\S}-4\pi^2\right).
\eeq Alternatively, the ringing frequency $\omega=\sqrt{-\sigma^2}$ is given by
\beq
\omega=\frac{2^{3/2}\pi}{3^{1/2}}\left(\frac{d}{\S}-4\pi^2\right)^{1/2},\label{omega_formula}
\eeq  which is in agreement with the numerically-determined predictions of both the FvK and Kirchhoff models.


\section{Kirchhoff equations}
The beam equations we have analysed are derived under the assumption that slopes are small, i.e.~$\vert w_x\vert\ll1$.  As we are primarily interested in shallow arches and the effect of stretchability, the linear equations capture the experimental observations very well.  Moreover, as we have shown, working in the linear regime enables us to obtain analytical formulas and key parameter values.  Next, for comparison, and in an attempt to obtain the most accurate prediction of dynamical growth rates possible, we model the problem using the nonlinear Kirchhoff equations.

We consider an extensible beam with centerline $\mathbf{r}(S)=x(S)\mathbf{e}_x+y(S)\mathbf{e}_y$ such that $S$ is the arclength in the reference (stress-free) configuration.   Letting $s$ denote the arclength in the current configuration, we define the axial stretch
\beq
\alpha=\alpha(S)=\frac{ds}{dS}
\eeq
To describe the geometry, it is convenient to define $\theta$ as the angle between the tangent to the centerline and the horizontal $x$-direction.  Then we have
\beq
\begin{split}
&\pd{x}{S}=\alpha\cos\theta\\
&\pd{y}{S}=\alpha\sin\theta,
\end{split}
\eeq
where prime denotes differentiation with respect to $S$.  To describe the mechanics, let $\mathbf{n}=n_x\mathbf{e}_x+n_y\mathbf{e}_y$ be the resultant force and $\mathbf{m}=m\mathbf{e_z}$ the resultant moment attached to the centerline by averaging the stress over the cross section.  Balancing linear and angular momentum yields  
\beq
\begin{split}
&\pd{n_x}{S}=\rho h\wid\pdd{x}{t}\\
&\pd{n_y}{S}-F\delta(S-S_c)=\rho h\wid\pdd{y}{t},\;\;\;\;x(S_c)=0\\
&\pd{m}{S}+\alpha(n_x\sin\theta-n_y\cos\theta)=0.
\end{split}
\eeq
Here the delta function signifies that the force is applied at the (unknown) material point $S_ c$ that corresponds to the fixed point $x=0$ in the lab frame.  To close the system, we must provide two constitutive laws and boundary conditions.  We assume a quadratic strain energy, so that the moment is linearly related to the curvature by
\beq
m=B\theta',
\eeq
where $B$ is the bending stiffness as defined in Equation \eqref{Bending}.  For an extensible beam, we also require a constitutive law relating the axial force to the stretch, which reads
\beq
n_x\cos\theta+n_y\sin\theta=Eh\wid(\alpha-1).
\label{Kirchhoff_constit}
\eeq
Note that $n_x\cos\theta+n_y\sin\theta$ is the tangential component of the force vector $\mathbf{n}$, thus \eqref{Kirchhoff_constit} essentially expresses Hooke's law.  The clamped boundary conditions are expressed as 
\beq
x(\pm L/2)=\pm L/2\mp\Delta L/2,\;\;\;\;y(\pm L/2)=0,\;\;\;\;\theta(\pm L/2)=0,
\eeq
and the fixed displacement condition reads $y(S_c)=y_0$, where $y_0$ is the fixed height at the point of the indentor (equal to $w_0$ in the FvK analysis).

We next non-dimensionalize the system.  We scale lengths by $L$, so that $S'=S/L$, $x'=x/L$, $y'=y/L$, the force $\mathbf{n}$ by $N=B/L^2$, moment $m$ by $M=B/L$, and as before we introduce the dimensionless force $f=F\times L^2/B$ and we again scale time by $t_*=L^2\left(\rho h\wid/B\right)^{1/2}$.  Inserting the scalings and dropping the primes, the system is 

\beq
\begin{split}
&\pd{x}{S}=\alpha\cos\theta,\;\;
\pd{y}{S}=\alpha\sin\theta\\
&\pd{n_x}{S}=\pdd{x}{t}\\
&\pd{n_y}{S}-f\delta(S-S_c)=\pdd{y}{t},\;\;\;\;x(S_c)=0,\;\;y(S_c)=y_0\\
&\pd{m}{S}+\alpha(n_x\sin\theta-n_y\cos\theta)=0\\
&m=\pd{\theta}{S}\\
&\S(n_x\cos\theta+n_y\sin\theta)=\alpha-1,
\end{split}
\label{Kirchhoff_full}
\eeq
where the stretching stiffness $\S=h^2/12L^2$ is as defined in the FvK analysis \eqref{eqn:Sdefn}.  The boundary conditions become
\beq
x(\pm 1/2)=\pm 1/2\mp d/2,\;\;\;\;y(\pm 1/2)=0,\;\;\;\;\theta(\pm 1/2)=0.
\eeq
%


\subsection{Statics}
To study the static deformation of the beam, we drop time derivatives in \eqref{Kirchhoff_full}.  We have solved the resulting system via a shooting method.  Starting at $S=-1/2$, we integrate the system forward with shooting variables 
\beq
\mathcal{V}=\{n_x,n_y(-1/2),m(-1/2),f,S_c\},
\eeq  
noting that the indentor force implies a jump in $n_y$ at $S=S_c$.  The conditions for a successful solution are $x(1/2)=(1-d)/2$, $y(1/2)=0$, $\theta(1/2)=0$, along with the conditions at the indentor $x(S_c)=0$, $y(S_c)=y_0$; the shooting variables $\mathcal{V}$ are iterated on until these five conditions are met.

\subsection{Upper bound on arch height}
\begin{figure}
\begin{centering}
\includegraphics[width=0.9\linewidth]{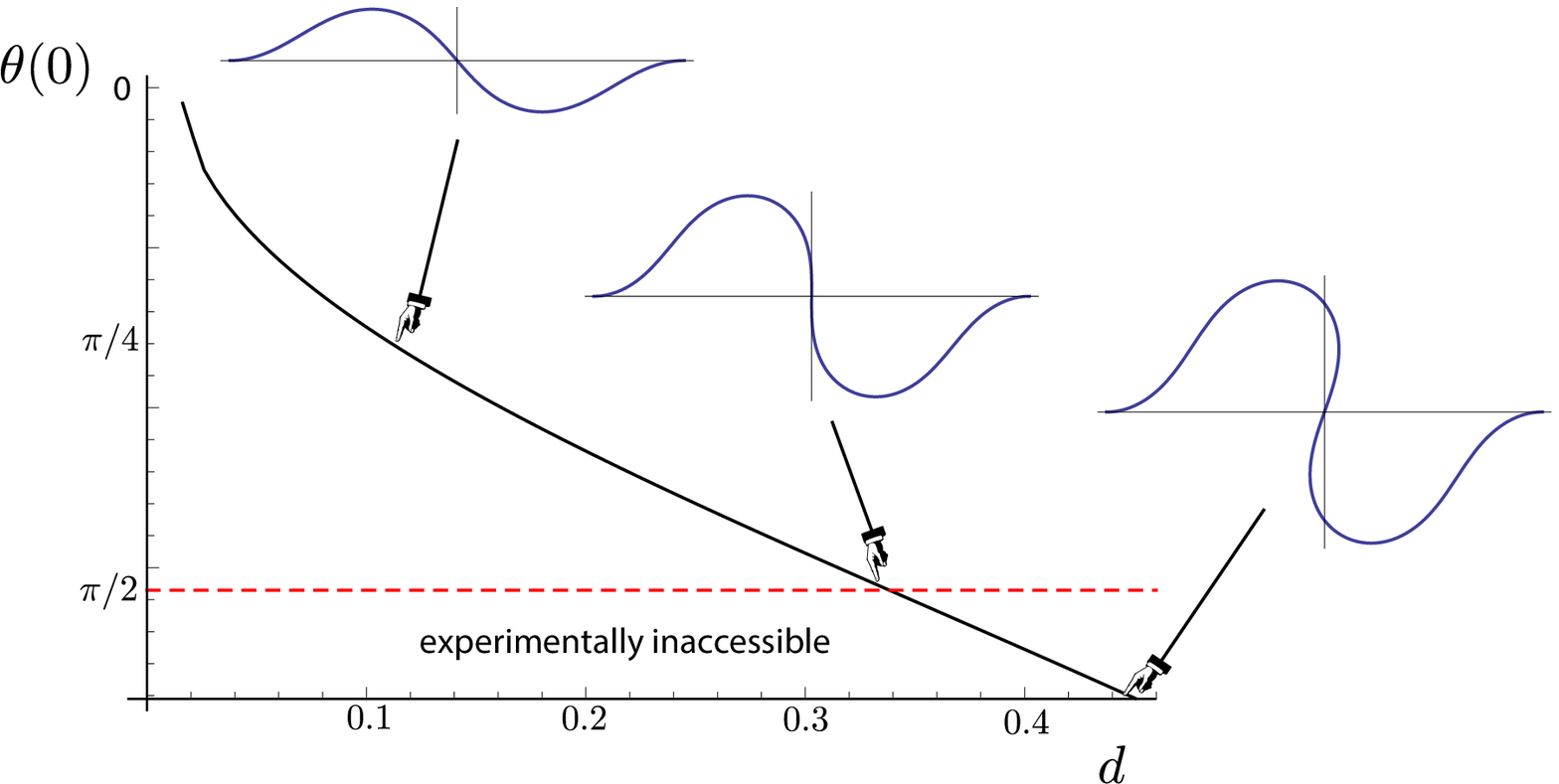}
\caption{The slope at the midpoint in the antisymmetric mode 2 solution plotted against the end-shortening $d$. Beyond $d_\star\approx0.34$, the beam shape becomes multi-valued and so is inaccessible experimentally.}\label{Theta_vs_d}
\end{centering}
\end{figure}

Using the Kirchhoff model, we can also determine a limit on the range of arch heights for which the indentation experiment is possible.  At a critical arch height (or critical end displacement $d_\star$), the beam is vertical at its midpoint when the indentor force and displacement $y_0$ vanish, i.e. when it reaches the antisymmetric mode 2 solution from which it subsequently snaps.  For $d>d_\star$, the beam reaches this point earlier in the deformation, and thereafter takes a multi-valued `S' shape in the antisymmetric state.  As the experimental indentor cannot maintain contact with a vertically sloping beam (or beyond), such deformations are experimentally inaccessible; $d_\star$ forms an upper limit on the end-shortenings that can be interrogated experimentally.  To find the critical height corresponding to $d_\star$, we need only focus on the antisymmetric mode 2 solution, and find the value of $d$ at which the slope $\theta(0)$ passes $\pi/2$.  Such a plot is given in Fig. \ref{Theta_vs_d}; we find the critical value $d_\star\approx0.34$.

\subsection{Dynamics}
To explore dynamics, we set $f=0$ and perform a linear stability analysis in the same manner as in Section \ref{Dyn_section}, but here we must add a perturbation to all variables; that is we expand $x=x_\nu+\epsilon x_pe^\sigma t$, $y=y_\nu+\epsilon y_pe^\sigma t$, etc.   The system at $O(\epsilon)$ is
\beq
\begin{split}
& x_p'=\alpha_p\cos\theta_\alpha-\alpha_\nu\theta_p\sin\theta_\alpha \\
& y_p'=\alpha_p\sin\theta_\alpha+\alpha_\nu\theta_p\cos\theta_\alpha \\
& n_{x_p}'=\sigma^2x_p\\
& n_{y_p}'=\sigma^2y_p\\
& \theta_p'=m_p \\
& m_p'=\alpha_p n_{x_\nu}\sin\theta_\alpha+\alpha_\nu n_{x_p}\sin\theta_\nu+\alpha_\nu n_{x_\nu}\theta_p\cos\theta_\nu\\
& \;\;\;\;\;\;\; - \alpha_p n_{y_\nu}\cos\theta_\nu-\alpha_\nu n_{y_p}\cos\theta_\nu+\alpha_\nu n_{y_\nu}\theta_p\sin\theta_\nu
\end{split}\label{OepsODE}
\eeq
with 
\beq
\alpha_p=\S\left(n_{x_p}\cos\theta_\nu-n_{x_\nu}\theta_p\sin\theta_\nu+n_{y_p}\sin\theta_\nu+n_{y_\nu}\theta_p\cos\theta_\nu\right).
\eeq
The boundary conditions at $O(\epsilon)$ are
\beq
x_p=y_p=\theta_p=0\;\;\;\;\text{at }S=\pm1/2.\label{OepsBC}
\eeq
To solve \eqref{OepsODE} - \eqref{OepsBC} for a given equilibrium solution $\{x_\nu,y_\nu,n_{x_\nu},\dots\}$, we implemented a determinant method, integrating three copies of the system from $S=-1/2$ with linearly independent initial conditions.  We then formulate the boundary conditions at $S=1/2$ as a determinant condition $\det M=0$, where $M=M(\sigma^2)$ is a matrix consisting of the boundary conditions \eqref{OepsBC} for each of the three copies.  If the system admits any solution with $\sigma^2>0$, the equilibrium solution is unstable and $\sigma$ defines the growth rate of the instability; if the determinant condition can only be satisfied for $\sigma^2<0$, the equilibrium solution is stable and the values $\omega=\sqrt{-\sigma^2}$ define the ringing frequencies.

\begin{figure}
\begin{centering}
\includegraphics[width=0.8\linewidth]{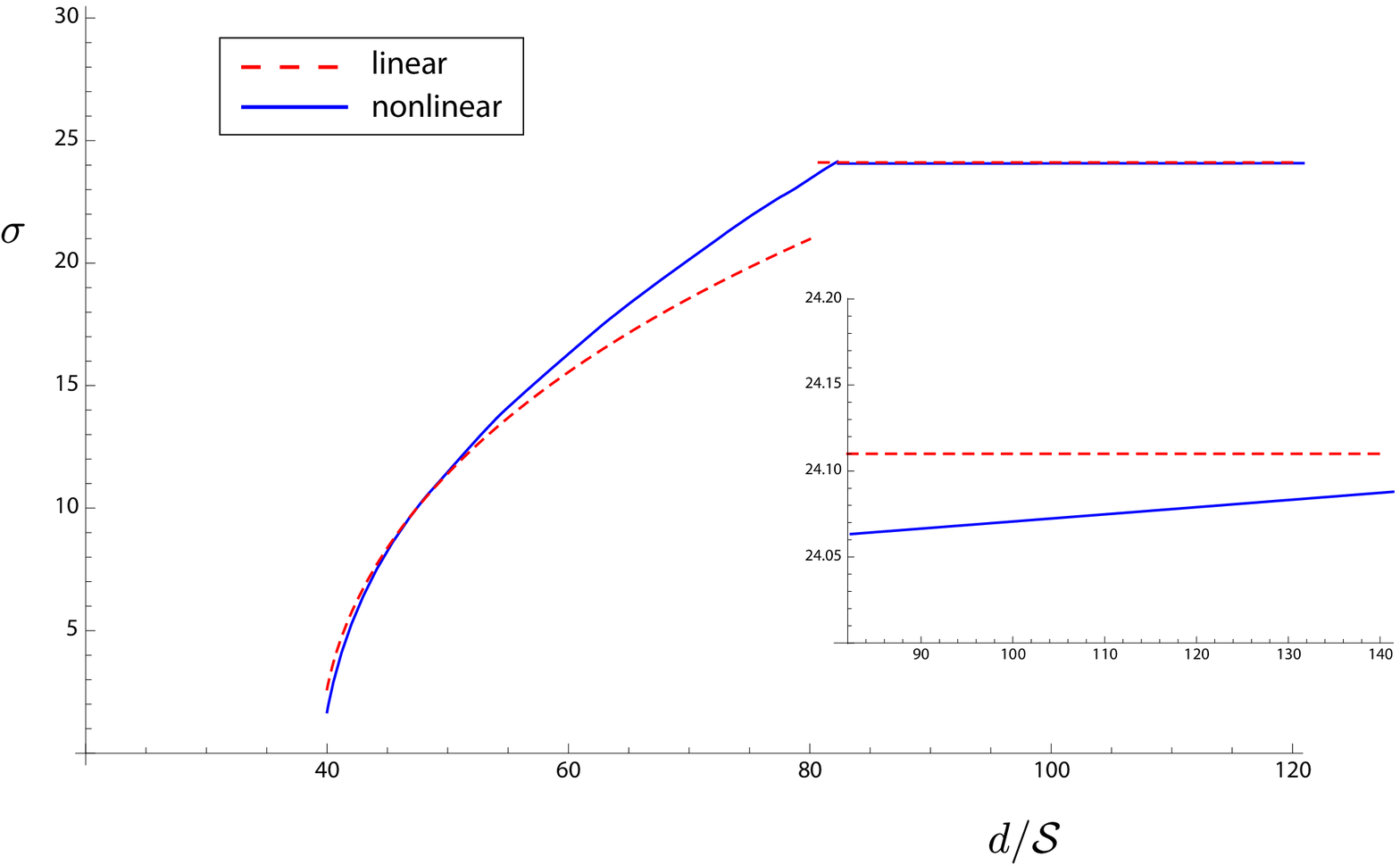}
\caption{Growth rate of snap-through instability for FvK (linear) and Kirchhoff (nonlinear) models.}\label{Sigma}
\end{centering}
\end{figure}
In Fig. \ref{Sigma}, we plot the growth rate $\sigma$ against $d/\S$ for both Kirchhoff and FvK models.   In the regime $d/\S<80.76$, the curved dashed line follows the approximation \eqref{sigma_approx}.  For $d/\S>80.76$ (inset), the Kirchhoff theory predicts a small increase in $\sigma$ with $d/\S$, but is well approximated by the FvK result $\sigma\approx24.11$.

\begin{figure}
\begin{centering}
\includegraphics[width=0.8\linewidth]{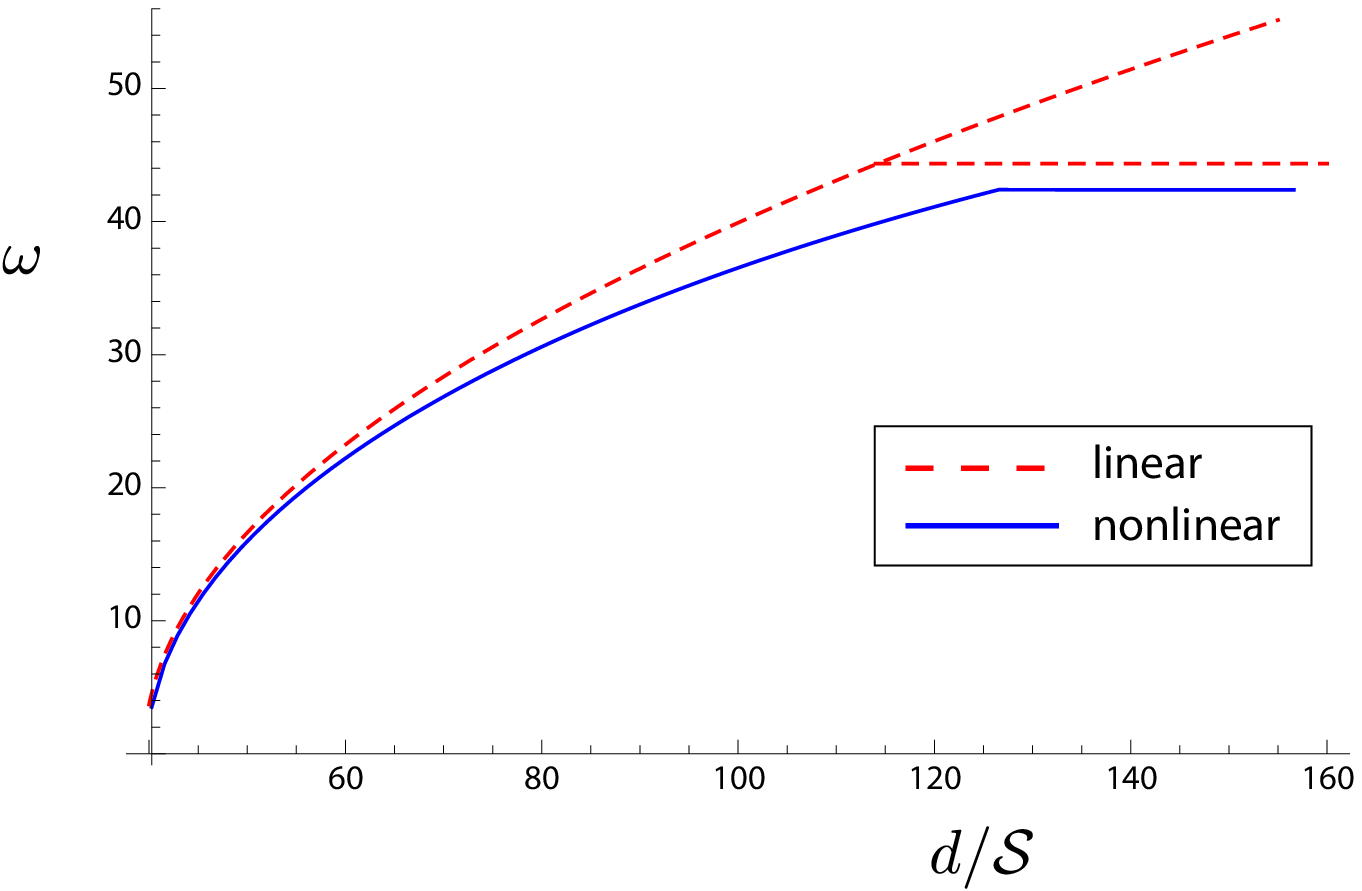}
\caption{Comparison of ringing frequency for FvK (linear) versus Kirchhoff (nonlinear).  }\label{Omega}
\end{centering}
\end{figure}

  Fig. \ref{Omega} gives a comparison of ringing frequencies for the two models.  The horizontal dashed line corresponds to the computed value $\omega\approx44.36$, while the curved dashed line is given by formula \eqref{omega_formula}. 

\section{Vibrations of a spherical shell}
In this section we outline the calculation of the vibration frequency of a spherical cap.  We consider a spherical cap with base diameter $L$, radius of curvature $R$, and thickness $h$.  (Note that the base diameter corresponds most closely to the width of the two-dimensional arches that we considered previously and so we use the same notation.) The normal displacement of the shell, denoted $w$, and the Airy stress function, denoted $\phi$, can be described by the Donnell-Mushtari-Vlasov equations, which read~\cite{soedel04}
\be{DMV}
\begin{split}
& B\Delta^2w+\frac{1}{R}\Delta\phi+\rho_sh\pdd{w}{t}=0\\
&\Delta^2\phi-\frac{Eh}{R}\Delta w=0.
\end{split}
\ee
Here $\rho_s$ is the density and $B=Eh^3/12(1-\nu^2)$ is the bending stiffness (with Poission ratio $\nu$ and Young's modulus $E$). The first of these equations expresses vertical force balance; the second expresses the compatibility of strains and allows us to deduce
\be{Airy}
\Delta\phi=\frac{Eh}{R}w.
\ee
Inserting \eqref{Airy} into the first equation of \eqref{DMV}, we obtain
\be{}
B\Delta^2w+\frac{Eh}{R^2}w+\rho_sh\pdd{w}{t}=0.
\ee

To find the vibration frequency of the popper, we perform a linear stability analysis assuming the shell remains radially symmetric.  Writing
$$w=w(r)e^{i\omega t},$$
where $r$ is the radial coordinate, we have
\be{DMV2}
B\Delta^2w+\left(\frac{Eh}{R^2}-\rho_sh\omega^2\right)w=0
\ee
where the symmetric Laplacian operator reads
$$\Delta w=w''(r)+\frac{1}{r}w'(r).$$
The popper is clear of the table when in the ringing state, thus we impose boundary conditions at the edge of the shell ($r=L/2$) corresponding to a free edge:
\be{BCPop}
\Delta w=\frac{d}{dr}\Delta w=0\quad\text{at }r=L/2.
\ee
Along with this, we have the following two conditions at the centre of the shell
\be{}
w'(0)=w'''(0)=0,
\ee
which ensure that no force is applied at the centre for the radially symmetric deformation.

To non-dimensionalize the problem, we scale lengths by the bending length 
$$ l_b=\left(\frac{BR^2}{Eh}\right)^{1/4}.$$
That is, we write $\tilde{w}=w/l_b$, $\tilde{r}=r/l_b$.  This natural length scale introduces the dimensionless frequency
$$\Omega=\left(\frac{\rho_sR^2}{E}\right)^{1/2}\omega=\frac{R}{c}\omega$$
where, as in the main paper, $c$ is the speed of sound in the material.  Inserting these scalings in \eqref{DMV2} and dropping tildes henceforth, the dimensionless problem is

\be{DMV3}
\begin{split}
&\Delta^2w+(1-\Omega^2)w=0\\
&w'(0)=w'''(0)=0\\
&\Delta w(a)=\frac{d}{dr}\Delta w(a)=0,
\end{split}
\ee
where $a=L/(2l_b)$.  We note that this eigenproblem is mathematically identical to the problem of the ringing of a circular disc with free edges, albeit with a shifted frequency. This is an example of the more general result of \cite{soedel1973} that such shell problems always reduce to the corresponding plate problem with the same boundary conditions and a modified frequency.

The general solution to \eqref{DMV3}, subject to the boundary conditions at $r=0$ (and a finite solution constraint at $r=0$) is
$$w(r)=AJ_0\left(r\left(\Omega^2-1\right)^{1/4}\right)+BI_0\left(r\left(\Omega^2-1\right)^{1/4}\right),$$
where $J_0$ and $I_0$ are Bessel functions of zeroth order, and $A$, $B$, are constants to be determined.

The linear stability analysis consists in finding the values of $\Omega$ for which $w$ has a non-trivial solution satisfying the boundary conditions at $r=a$, \eqref{BCPop}.  This leads to the determinant condition $\det(M)=0$ where
\be{}
M(\Omega)=\left(\begin{matrix}
I_0\left(a\left(\Omega^2-1\right)^{1/4}\right) & -J_0\left(a\left(\Omega^2-1\right)^{1/4}\right) \\
I_1\left(a\left(\Omega^2-1\right)^{1/4}\right) & J_1\left(a\left(\Omega^2-1\right)^{1/4}\right)
\end{matrix}\right).
\ee
We find numerically that the smallest solution of this is given by
$$a\left(\Omega^2-1\right)^{1/4}\approx3.196=:\lambda$$
Inserting this into the scalings, and recalling the characteristic timescale $t_*=2\sqrt{3}\frac{L^2}{ch}$ defined in section \ref{SetupSec} and used in the main paper, we obtain the following expression for the dimensional ringing frequency:
\be{}
\omega^2=\frac{c^2}{R^2}+16\frac{\lambda^4}{1-\nu^2}\frac{1}{t_*^2}.
\ee
From this, the ringing period $t_r:=2\pi/\omega$ can be expressed as

\be{}
t_r=\frac{\pi\sqrt{1-\nu^2}}{2\lambda^2}t_*(1+\epsilon)^{-1/2},
\ee
where we anticipate that 
$$\epsilon:=\frac{3}{4}\frac{1-\nu^2}{\lambda^4}\frac{\Lbase^4}{R^2h^2}\ll1.$$
Indeed, for the typical popper experiments presented here, $L=R$, $R/h=5$ and $\nu\approx0.5$, which gives $\epsilon=0.134$, while for the jumping disk $\epsilon=0.003$.  If we neglect the $\epsilon$ term, we obtain the universal shell relation

$$t_r\approx\frac{\pi\sqrt{1-\nu^2}}{2\lambda^2}t_*\approx0.133t_*.$$

\appendix{The general solution of the static problem}

The general solution of \eqref{eq:linbeam} may be written
\beq
w(x)=\begin{cases}
A_-+B_-x+C_-\cos\tau x+D_-\sin\tau x,\quad -1/2<x<0\\
A_++B_+x+C_+\cos\tau x+D_+\sin\tau x,\quad 0<x<1/2.
\end{cases}
\label{linbeam:gensoln}
\eeq
The continuity of $w''$ at $x=0$ gives immediately that $C_-=C_+=C$. In turn, the continuity of $w$ at $x=0$ gives that $A_-=A_+=A$ and, further, that
\beq
A+C=\wo.
\eeq The clamped boundary conditions give
\beq
0=A\pm B_\pm/2+C\cos\tau/2 \pm D_\pm \sin\tau/2
\eeq and
\beq
0=B_\pm\mp C\tau\sin\tau/2 +D_\pm\tau \cos\tau/2.
\eeq Hence
\beq
(B_++B_-)/2+(D_++D_-)\sin\tau/2=0
\eeq and
\beq
B_++B_-+(D_++D_-)\tau\cos\tau/2=0
\eeq so that either
\beq
D_++D_-=0,\quad B_++B_-=0
\label{solvability:symm}
\eeq or
\beq
\tan\tau/2=\tau/2.
\label{solvability:asymm}
\eeq Note that the first possibility corresponds to a symmetric mode of deformation since  $B$ and $D$ multiply the odd terms in $w(x)$. The latter possibility corresponds to an asymmetric mode (but \emph{not}  antisymmetric). We need to consider each of these possibilities separately.

\subsection{The symmetric mode}

In the symmetric mode we satisfy \eqref{solvability:symm} by letting
\beq
D_+=-D_-=D,\quad B_+=-B_-=B.
\eeq It is then a simple matter to see that \eqref{linbeam:gensoln} is the symmetric shape
\beq
w(x)=A+B|x|+C\cos\tau x+D\sin\tau |x|,\quad -1/2<x<1/2.
\eeq Using the imposed displacement and slope conditions at $x=0$ along with the clamped conditions at $x=1/2$ we find that
\begin{eqnarray}
A&=&\wo\frac{1-\cos\tau/2-\tau/2\sin\tau/2}{2-2\cos\tau/2-\tau/2\sin\tau/2}\\
B&=&\wo\frac{\tau\sin\tau/2}{2-2\cos\tau/2-\tau/2\sin\tau/2}\\
C&=&\wo\frac{1-\cos\tau/2}{2-2\cos\tau/2-\tau/2\sin\tau/2}\\
D&=&-\wo\frac{\sin\tau/2}{2-2\cos\tau/2-\tau/2\sin\tau/2}.
\end{eqnarray}

The shape is thus determined, although we currently do not have any indication of the value of $\tau$. To determine this quantity, we need to make use of the given end-end displacement, $d$. By definition we have that
\beq
d-\tau^2\S=\int_{-1/2}^{1/2}\tfrac{1}{2}w_x^2~\upd x=\int_{0}^{1/2}w_x^2~\upd x=\wo^2\frac{\tau\left(2\tau+\tau\cos\tau/2-6\sin\tau/2\right)}{\left(\tau\cos\tau/4-4\sin\tau/4\right)^2}.
\eeq This shows that for a given value of $d$, $\wo$ may be written parametrically in terms of $\tau$ as
\beq
\wo^2=(d-\tau^2\S)\frac{\left(\tau\cos\tau/4-4\sin\tau/4\right)^2}{\tau\left(2\tau+\tau\cos\tau/2-6\sin\tau/2\right)}.
\label{eqn:w0symm}
\eeq 

Before any indentation occurs, the shape is given by $w=\tfrac{\wo}{2}\left(1+\cos2\pi x\right)$, which corresponds to $\tau=2\pi$ giving $\wo=2(d-4\pi^2\S)^{1/2}/\pi$. Also, the right hand side of \eqref{eqn:w0symm} is a decreasing function of $\tau\geq0$ so that, as indentation progresses (i.e.~as $\wo$ decreases) $\tau$ must increase. We therefore expect that the value of $\tau$ should increase up until the point at which either $\tau/4=\tan\tau/4$ (so that $\tau\approx17.97$) or until $\tau=(d/\S)^{1/2}$. In either one of these cases $\wo=0$. However, since the asymmetric mode has $\tau=\taus\approx8.99$ (the solution of $\taus/2=\tan\taus/2$) there is also the possibility that the symmetric solution is replaced by the asymmetric one at a displacement
\beq
\woc\approx 0.62(d-80.76\S)^{1/2}.
\eeq 

Now, in the symmetric mode the indentation force is given by
\beq
f=[w''']^+_-=-2\tau^3D=2\wo\frac{\tau^3\sin\tau/2}{2-2\cos\tau/2-\tau/2\sin\tau/2}.
\label{eqn:symmforcelaw}
\eeq Because of the nonlinear relationship between the displacement $\wo$ and $\tau$ in \eqref{eqn:w0symm}, the force law \eqref{eqn:symmforcelaw} is, in general, a nonlinear function of the displacement $\wo$. However, we note that at the displacement $\woc$ the force is given by $f^{(c)}\approx -207.75 \woc$.

\subsection{The asymmetric mode}

In the asymmetric mode, we have that the compression $\tau=\taus\approx8.99$ where $\taus$ is the smallest solution of the equation $\taus/2=\tan(\taus/2)$. Returning to the equations for the unknown coefficients and replacing $\sin(\taus/2)$ by $\cos(\taus/2)\taus/2$ we have
\begin{eqnarray*}
A+C&=&\wo\\
A\pm B_\pm/2+C\cos\taus/2 \pm D_\pm\taus/2 \cos\taus/2&=&0\\
B_\pm\mp C\taus^2/2\cos\taus/2 +D_\pm\taus \cos\taus/2&=&0\\
(B_++B_-)+(D_++D_-)\taus\cos\taus/2&=&0,
\end{eqnarray*} noting that the third of these equations implies the fourth. We find
\beq
A+C(1+\taus^2/4)\cos\taus/2=0
\eeq and so
\beq
A=-\wo\frac{(1+\taus^2/4)\cos\taus/2}{1-(1+\taus^2/4)\cos\taus/2}, \quad C=\frac{\wo}{1-(1+\taus^2/4)\cos\taus/2}.
\eeq
%
%
The requirement that the first derivative is continuous at the indentation point implies
\beq
B_-+D_-\tau=B_++D_+\tau.
\label{asymm:BD}
\eeq Since we have already used the continuity of $w''$ at $x=0$ to determine $C$ the only remaining relationship comes from the imposed confinement, $d$. As it is clear that the coefficients $B_\pm,D_\pm\propto \wo$, it is convenient to write
\beq
\frac{d-\taus^2\S}{\wo^2}=\frac{1}{2}\int_{-1/2}^{1/2}\omega_x^2~\id x
\label{asymm:dlcond}
\eeq where $\omega=w/\wo$. In this expression, 
it is possible to express $D_+/\wo$, $B_\pm/\wo$ in terms of $D_-/\wo$ and thereby determine these coefficients as $\wo$ varies for a given value of $d$ by solving \eqref{asymm:dlcond} numerically. This is useful for determining the predicted shape of the beam. However, our main interest lies in understanding the force as a function of displacement and, possibly the bending energy of a given deformation. In this regard, we have that the indentation force is given by
\beq
f=[w''']_-^+=-\taus^3(D_+-D_-).
\eeq  The quantity of most interest then is $D_+-D_-$, which can be written
\beq
D_+-D_-=-C\frac{\taus\cos\taus/2}{1-\cos\taus/2}.
\eeq Hence
\beq
f=\alpha(\taus)\wo
\eeq where $$\alpha(\taus)=\frac{\taus^4\cos\taus/2}{2-(2+\taus^2/4)\cos\taus/2}\approx-207.75$$ is a constant, independent of the value of $d$. We note that the force in this regime is therefore linearly proportional to $\wo$. We see that this force decreases as the amount of indentation approaches $\wo=0$ (i.e.~as the bump approaches the second Euler buckling mode).

\subsection{Transition from symmetric to asymmetric modes}

To determine when the transition from symmetric to asymmetric occurs, we compare the energies of the two different configurations. We note that, by construction, the imposed compression, $d$, is the same in both configurations and so the stretching energy must be constant. To simplify the calculation further, we note that integrating \eqref{eqn:constraint} by parts and using \eqref{eqn:beameqnnd} gives
\beq
2(d-\S \tau^2)=\int_{-1/2}^{1/2}w_x^2~\upd x=\tau^{-2}\int_{-1/2}^{1/2}~[w_{4x}-f\delta(x)]w~\upd x
\eeq so that
\beq
2\tau^2(d-\S\tau^2)=-fw_0+\int_{-1/2}^{1/2}w_{xx}^2~\upd x=-f\wo+2U_B
\eeq and thus
\beq
U_B=\wo f/2+\tau^2 (d-\S\tau^2).
\label{eqn:bendingenergy}
\eeq 

This relationship holds whichever mode (symmetric or asymmetric) the system is in. However, where both modes exist for a given value of $\wo$, the asymmetric mode has both a lower force $f$ and a lower value of the compression $\tau$ than the symmetric mode. It is therefore clear from \eqref{eqn:bendingenergy} that, when the asymmetric mode exists, it has a lower bending energy (and so should be expected to be observed in preference to) the symmetric mode.

\bibliographystyle{jfm}